\documentclass[11pt,a4paper]{article}
\usepackage{hyperref}
\usepackage[super,sort,compress]{cite}

\usepackage{graphicx}% Include figure files
\usepackage{dcolumn}% Align table columns on decimal point
\usepackage{bm}% bold math
\usepackage[usenames,dvipsnames]{xcolor} 
\usepackage[utf8]{inputenc}
\usepackage{latexsym}
\usepackage{amsfonts}
\usepackage{amssymb}
\expandafter\let\csname equation*\endcsname\relax % put these lines before \usepackage{amsmath} otherwise it gives an error
\expandafter\let\csname endequation*\endcsname\relax
\usepackage{amsmath}
\usepackage{psfrag}
\usepackage{rotating}
\usepackage{pgf,pgfsys,pgffor} %
\usepackage{pgfplots}
\usepackage{pgfplotstable}
\usepackage{tikz}
\usetikzlibrary{intersections,arrows,decorations.pathmorphing,backgrounds,fit,mindmap,calc,through,scopes,fadings,positioning,automata,calendar,shapes,er,matrix,folding,patterns,petri,plothandlers,plotmarks,shadows,topaths,through,trees,pgfplots.units,decorations}  
\usepgfplotslibrary{fillbetween}
\usepackage{subfigure}

\usepackage{multirow}

\usepackage{bm}
\newcommand{\ave}[1]{\left \langle {#1} \right \rangle}	% expectation
\newcommand{\ee}{e} % for exponential		
\newcommand{\dd}{\textrm{ d}} % for differential term
\newcommand{\ii}{{i\mkern1mu}} % for imaginary unit
\newcommand{\beq}{\begin{equation}}
\newcommand{\eeq}{\end{equation}}
\newcommand{\So}{S} % original spectrum
\newcommand{\Fo}{\widetilde{F}} % base for original spectrum
\newcommand{\ao}{\widetilde{a}} % base for original spectrum
\newcommand{\Sr}{\hat{S}} % original spectrum

\usepackage{authblk}
\newcommand{\refcite}[1]{\cite{#1}}

\begin{document}

%%%%%%%%%%%%%%%%%%%%% Publisher's Area please ignore %%%%%%%%%%%%%%
%\catchline{}{}{}{}{}
%%%%%%%%%%%%%%%%%%%%%%%%%%%%%%%%%%%%%%%%%%%%%%%%%%%%%%%%%%%%%%%%%%%

\title{\bf ROLE OF THE FILTER FUNCTIONS IN NOISE SPECTROSCOPY}

% \title{\bf IMPROVED NOISE SPECTROSCOPY BY FILTER FUNCTION DESIGN AND NON-NEGATIVE LEAST SQUARES APPROACH}

%\author{Nicola Dalla Pozza\footnote{Electronic address: nicola.dallapozza@unifi.it (Corresponding author)}, Stefano Gherardini\footnote{Electronic address: stefano.gherardini@unifi.it}, Matthias M. M{\"u}ller\footnote{Electronic address: ma.mueller@fz-juelich.de}, Filippo Caruso\footnote{Electronic address: filippo.caruso@unifi.it}}
%
%\address{$\phantom{}^\ast \phantom{}^\dagger \phantom{}^\mathsection$ Dept. of Physics and Astronomy \& LENS, University of Florence\\
%Via Carrara 1, I-50019 Sesto Fiorentino, Italy. \\
%$\phantom{}^\ddag$Institute of Quantum Control, Peter Gr\"{u}nberg Institut\\
%Forschungszentrum J\"{u}lich, J\"{u}lich, Germany.}

\author{Nicola Dalla Pozza\footnote{Electronic address: nicola.dallapozza@unifi.it (Corresponding author)}, Stefano Gherardini\footnote{Electronic address: stefano.gherardini@unifi.it}, Matthias M. M{\"u}ller\footnote{Electronic address: ma.mueller@fz-juelich.de}, Filippo Caruso\footnote{Electronic address: filippo.caruso@unifi.it}}

\affil{$^\ast$$^\dagger$$^\mathsection$ Dept. of Physics and Astronomy \& LENS, University of Florence\\
Via Carrara 1, I-50019 Sesto Fiorentino, Italy.\\
$^\ddag$ Institute of Quantum Control, Peter Gr\"{u}nberg Institut\\
Forschungszentrum J\"{u}lich, J\"{u}lich, Germany.}

\maketitle

%\begin{history}
%\received{Day Month Year}
%\revised{Day Month Year}
%\accepted{Day Month Year}
%\comby{(xxxxxxxxxx)}
%\end{history}

\begin{abstract}
%\ndp{COMMENTO TITOLO: Nel paper usiamo "filter function design" e il relativo acronimo FFD, ma mi sembra troppo generico, soprattutto se in futuro vogliamo citare questo metodo o usare altri metodi di filter function design. Che ne dite di Bandwidth Overlap Design (BOD) of filter functions?}
The success of quantum noise sensing methods depends on the optimal interplay between properly designed control pulses and statistically informative measurement data on a specific quantum-probe observable.
To enhance the information content of the data and reduce as much as possible the number of measurements on the probe, the filter orthogonalization method has been recently introduced. The latter is able to transform the control filter functions on an orthogonal basis allowing for the optimal reconstruction of the noise power spectral density. 
In this paper, we formalize this method within the standard formalism of minimum mean squared error estimation and we show the equivalence between the solutions of the two approaches.
Then, we introduce a non-negative least squares formulation that ensures the non-negativeness of the estimated noise spectral density. Moreover, we also propose a novel protocol for the design in the frequency domain of the set of filter functions. The frequency-designed filter functions and the non-negative least squares reconstruction are numerically tested on noise spectra with multiple components and as a function of the estimation parameters. 
%The abstract should summarize the context, content and conclusions of the paper in less than 200 words. It should not contain any references or displayed equations. Typeset the abstract in 8 pt Times roman with baselineskip of 10~pt, making an indentation of 1.5 pica on the left and right margins.
\end{abstract}

\section{\label{sec:introduction}Introduction}

The sensing of stochastic fluctuations entering in the dynamics of an open quantum system is a crucial issue to be addressed to make quantum technologies operating and high-performing.\cite{HallPRL2009,Cole2009,Degen2017}
Stochastic fluctuations usually originate from external noise sources, and, in this case, one has to refer to quantum noise sensing techniques.\cite{Yuge2011,Alvarez2011,Bylander2011,Paz-Silva2014,Mueller2016b,Norris2016,FreyNatComm2017,Mueller2018,HernandezPRB2018,Do2019} Also quantum noise, i.e. the inevitable decoherence of a quantum system through entanglement with its quantum environment, can very often be modeled by a classical stochastic field \cite{Gu2019}.

The aim of noise sensing is to infer spectral properties of noise fluctuation fields affecting a quantum system used as a probe. Usually, the most significant quantity to reconstruct is the noise \emph{power spectral density} $\So(\omega)$, which is formally defined as the Fourier transform of the noise auto-correlation function. In order to selectively measure specific portions of the noise power spectral density, the system is coherently manipulated with properly designed control signals.\cite{UhrigNJP2008,Biercuk2011,KotlerPRL2013,HaberlePRL2013,Paz-Silva2014,NobauerPRL2015,Mueller2018} In the context of noise sensing, the squared module of each control signal Fourier transform is called \emph{filter function} $F(\omega)$. Its introduction has become important since it was shown that $\So(\omega)$ and $F(\omega)$ are linked by a  general relation\cite{KofmanPRL2001,GordonJPB2007,Zwick2016} quantifying the overlap $\chi$ between the effects of noise and control on the quantum probe dynamics, i.e.
\beq
\int_{-\infty}^{+\infty}\So(\omega) F(\omega) \dd \omega = \chi\ ,
\label{universalRelation}
\eeq
irrespective of the physical realization of the control pulses. In the sensing protocol, we manipulate the system with different control signals, and from the statistics of the data we aim at inverting the relation \eqref{universalRelation}.

Despite this universal formula, the design of filter functions can fail for the purpose of sensing once applied to the quantum sensor. The reasons of this can be bi-fold\cite{MuellerSensingZeno2019}: (i) The power spectral density we want to infer is completely unknown a-priori; (ii) the physical implementation of the control pulses is limited by experimental constraints and imperfections. To overcome these issues, the \emph{Filter Orthogonalization} (FO) method has been recently proposed\cite{Mueller2018} with the aim to make the sensing procedure \emph{robust}, irrespective of the presence of imperfections in control and detection apparata, respectively. The motivation behind this method is to successfully use also the information coming from filter functions that have not been designed as an orthogonal basis due to physical limitations. In this respect, the FO protocol determines the basis transformation that orthogonalizes a set of
% renders designed 
filter functions 
% orthogonal 
to solve the inference problem of noise spectral properties on a completely informative (mathematical) working space, leading to a high reconstruction fidelity. The method has been lately experimentally adopted on an engineered Bose-Einstein condensate of $^{87}$Rb atoms realized on an atom chip.\cite{Do2019} It is worth noting that a similar approach, also relying on the use of the Gramian matrix, has been lately proposed to facilitate the derivation of the quantum Fisher information in many quantum statistical models.\cite{GenoniArxiv2019}

In this paper, we formalise the FO method within the framework of minimum mean squared error (MSE) estimation.\cite{Book-Luenberger-69,Book-Scharf} In particular, we show that the FO estimation is obtained as a linear combination of the filter functions, where the coefficients are given by the solution of the MSE minimization by means of Least Squares (LS). We further analyze this minimization and address the issue that the reconstructed power spectral density may exhibit negative values at some frequencies after an estimation from noisy data. We propose a Non-Negative Least Squares (NNLS) minimization of the MSE, which can be solved numerically. 

Since the FO estimate is given by a linear combination of the filter functions within a range of frequencies, one could improve the design of each control pulse by also ensuring the estimation procedure fulfills specific properties in the frequency domain. In particular,
%The fact that the FO estimate is a linear combination of the filter functions allows to understand the spectrum frequencies that can be recovered with this approach. This insight suggests that, 
to obtain a good spectral density estimation on a given frequency interval, the filter functions should cover it uniformly, possibly with disjoint frequency support. We develop this intuition into a novel strategy for the design of the set of filter functions, which are chosen to have a fixed ratio of the bandwidth overlap of their main peaks. This Bandwidth-Overlap Design (BOD) is then tested in comparison with two sensing protocols employing multipulse sequences, i.e. Periodic Dynamical Decoupling (PDD) sequences\cite{Khodjasteh2005,Alvarez2011,Degen2017} and Carr-Purcell  (CP) sequences.\cite{Carr1954,Alvarez2011} 

The paper is organized as follows. In Section \ref{sec:problem} we introduce the MSE minimization and from its solution we recover the FO estimation. We cover the minimization by means of least squares and non-negative least squares. In Section \ref{sec:design} we introduce the bandwidth-overlap design of filter functions. In Section \ref{sec:numerical} we numerically compare the performance of this design against PDD and CP protocols, both with least squares and non-negative least squares estimation. Section \ref{sec:discussion} concludes the paper.
%\mmm{In in Section \ref{sec:problem}, we formalize the estimation problem and we provide a new solution to it.}
%\textcolor{red}{[Add details on what we are going to prove/show]}.

\section{\label{sec:problem}Problem Statement and Results}

Let us consider a wide-sense stationary stochastic process $\Omega(t)$, that is, a stochastic process that is stationary in the mean, $\ave{\Omega(t)} = m_{\Omega}$, and also in its auto-correlation, $\ave{\Omega(t) \Omega(t-\tau)}=g(\tau)$. This stochastic process could be for instance a time fluctuating classical field that couples a quantum system to an external environment.\cite{Mueller2016b,Mueller2018, Do2019} Without loss of generality, we assume the stochastic process to have zero mean, i.e., $m_{\Omega}=0$.

For a wide-sense stationary process, the power spectral density is well-defined from the Fourier transform of the auto-correlation function,
\beq
\So(\omega) = \int_{-\infty}^{+\infty} g(t) \ee^{-\ii \omega t} \dd t \ ,
\label{def:spectrum}
\eeq
which can also be inverted to recover $g(t)$ as
\beq
g(t) = \frac{1}{2\pi}\int_{-\infty}^{+\infty} \So(\omega) \ee^{\ii \omega t} \dd \omega \ .
\label{def:auto-correlation}
\eeq

In many instances, the quantum system can be coherently manipulated via a control signal $\Omega_c(t)$, which acts on each single stochastic realization of $\Omega(t)$ as a modulation. After this interaction, at the end of the applied control protocol, a measurement is usually performed. Then, from the statistics of repeated final measurements, one can relate the true power spectral density $\So(\omega)$ with the action of the control pulse by introducing the second-order correlation function $\chi$: 
\beq
\chi = \int_{-\infty}^{+\infty}\int_{-\infty}^{+\infty} \Omega_c(t) \Omega_c(t') \langle \Omega(t) \Omega(t') \rangle \dd t \dd t' = \int_{-\infty}^{+\infty} \So(\omega) F(\omega) \dd \omega \ .
\label{def:chi}
\eeq
In Eq.\,\eqref{def:chi}, $F(\omega)$ denotes the filter function, which is defined as
%\ndp{the proper definition of filter function is $F(\omega)$ or $\Omega_c(t)$?} \mmm{MM: the (frequency) filter function is $F(\omega)$.. $\Omega_c(t)$ is the control field or pulse modulation function as it is called in the coherent (DD) approach. The definition should be fine.}
\beq
F(\omega) = \frac{1}{2\pi} \left \vert \int_{-\infty}^{+\infty} \Omega_c(t)\,\ee^{-\ii \omega t} \dd t \right \vert^2.
\eeq
This overlap between filter function and noise spectral density is also called the universal formula of quantum sensing.\cite{KofmanPRL2001,GordonJPB2007} This is the case for instance of 
%Just as an example to better understand the physical meaning of $\Omega_c$ and %$\Omega$, the reader can refer - where suitable - to
%the formulation and experimental details 
the experimental results in Ref.~\refcite{Do2019}\,, where the control pulse $\Omega_c$ is a resonant microwave field with amplitude modulated by a periodic square-wave, which drives the transition between the ground and excited state of a two-level quantum sensor, engineered within a Rubidium Bose-Einstein condensate. In addition to the control field, the sensor is also driven by some external noise which we describe as a semi-classical stochastic process $\Omega(t)$. In the specific case of Ref.~\refcite{Do2019}\,, also a sequence of projective measurement in the so-called \emph{weak Zeno regime}\cite{MuellerAnnalen2017,GherardiniQST2017} is applied on the probe, with a rate proportional to the periodicity of $\Omega_c$. This leads to the advantage to allow for the inference of $\chi$ by directly measuring the atomic population in the sensor ground state.
%\ndp{spiegare Do2019\cite{Do2019} in questi termini dando l'interpretazione fisica di cosa sono $\Omega$, $\Omega_c$, e come si ricava $\chi$.} \mmm{MM: or should we rather state it in terms of the pulse modulation function $y(t)$ as in ther formulation that we have used in \cite{Mueller2018} and that is taken from \cite{Degen2017}? But anyways, the main equation $\chi=\int S F d\omega$ is the same.}

The aim of many sensing techniques is to provide an estimation $\Sr(\omega)$ of the true spectral density $\So(\omega)$. A set of filter functions $F_n(\omega)$ is employed in the sensing protocol, each of them using a different control signal $\Omega_c^{(n)}(t)$, with $n$ being an index representing a configuration of its parameters. We estimate $\So(\omega)$ by exploiting the relation between the power spectral density, the filter functions and the corresponding measurement data $\chi_n$,
\beq
\int_{-\infty}^{+\infty} \So(\omega) F_n(\omega) \dd\omega = \chi_n \ .
\label{chiN}
\eeq
A simple idea to estimate $\So(\omega)$ consists in discretizing both $\So(\omega)$ and $F_n(\omega)$ in the frequency domain, defining the column ${\bf \So}_k = \So(k \Delta \omega)$ and the matrix ${\bf F}_{n,k} = F_n(k \Delta \omega)$ and solving the linear system
\beq
{\bf F} \, {\bf \So} = {\boldsymbol \chi} \, ,
\label{def:ls_system}
\eeq
with ${\boldsymbol \chi}$ being the vector collecting the second-order correlation functions $\chi_n$.
%function, ${\boldsymbol \chi}_n = \chi_n$. 
It is easy to realize that this approach quickly fails when the frequency discretization of ${\bf F}$, obtained for instance from a \emph{discrete} Fourier transform of the corresponding control, gives a number of unknowns (or degrees of freedom) in ${\bf \So}$ usually orders of magnitude higher than the number of controls (or measurements)
%, i.e. system constraints, 
that one usually has available. Attempts to solve the system with Moore-Penrose pseudoinverse usually results in bad estimates due to ill-conditioning of the inversion problem. This approach was investigated in Ref.~\refcite{Alvarez2011} where a stable inverse has been achieved by choosing the main peaks of the filter functions as the discrete values of $\omega$ (leading to a square matrix ${\bf F}$).
Furthermore, the system \eqref{def:ls_system} gives an idea of what a good filter function would be. If each filter function excites a narrow band of frequencies of $\So(\omega)$, i.e. one or few of the components ${\bf \So}_k$, the inversion would be simplified. We will explore this intuition in Section \ref{sec:design} to design improved filter functions.

Rather than directly solving the system \eqref{def:ls_system}, let's define our spectrum estimate as a finite linear combination of the filter functions, i.e.,
\beq
\Sr(\omega) = \sum_{n=1}^N a_n F_n(\omega) \ ,
\eeq

We then define the optimal estimation as the solution of a minimization problem with the mean squared error criterion,\cite{Book-Luenberger-69}
\beq
%\min_{\bf a} MSE, \quad  
{\rm MSE} = \int_{-\infty}^{+\infty} \left [ \So(\omega) - \Sr(\omega)  \right ]^2 \dd \omega 
\label{def:MSE}
\eeq
with respect to the vector ${\bf a}$ collecting the coefficients $[{\bf a}]_n=a_n$.

Here, it is worth noting that in the space of the control signals defined on the Fourier domain\footnote{For a more rigorous definition of the space of signals refer to signal processing textbooks such as  Ref.~\refcite{KennedySadeghi}.}, the set of filter function $F_n$ is in general not orthonormal, in the sense that
\beq
\int_{-\infty}^{+\infty} F_n(\omega) F_m(\omega) \dd\omega = G_{n,m}\neq \delta_{n,m} \ , 
\label{def:gramian}
\eeq
where $\delta_{n,m}$ denotes the Kronecker delta. 
%As previously explained in Sec.\,\ref{sec:problem}, 
The non-orthonormality of the filter functions usually comes from physical limitations in the implementation of the control pulses rather than being designed on purpose. All the inner products $G_{n,m}$ are collected in the Gramian matrix ${\bf G}$. If we introduce an orthonormal basis $\{\Fo\}_{k=1}^{K}$ for $F_n$, we can write  
\beq
F_n(\omega) = \sum_{k=1}^K b_{n,k} \Fo_k(\omega) \ ,
\eeq
where ${\bf B}$ is the matrix collecting the coefficients $b_{n,k}=[{\bf B}]_{n,k}$, verifying ${\bf G}={\bf B}{\bf B}^T$. Thus, if we write $\So(\omega)$ in the orthonormal basis $\{\Fo\}_{k=1}^{K}$ that we formally extend for indices $k \geq K+1$, i.e.,
\beq
\So(\omega) = \sum_{k=1}^{+\infty} c_{k} \Fo_k(\omega)
\eeq
then we can rewrite the mean squared error as
\begin{align}
{\rm MSE} & = \int_{-\infty}^{+\infty} \left [ \sum_{k=1}^{+\infty} c_{k} \Fo_k(\omega) - \sum_{n=1}^N a_n \sum_{k=1}^K b_{n,k} \Fo_k(\omega) \right ]^2 \dd \omega \\
& = \int_{-\infty}^{+\infty} \left [ \sum_{k=1}^K \left ( c_{k}  - \sum_{n=1}^N a_n b_{n,k} \right ) \Fo_k(\omega) + \sum_{k=K+1}^{+\infty} c_{k} \Fo_k(\omega)\right ]^2 \dd \omega \\
& = \sum_{k=1}^K \left ( c_{k}  - \sum_{n=1}^N a_n b_{n,k} \right )^2 + \sum_{k=K+1}^{+\infty} c_{k}^2 \\ 
& = \sum_{k=1}^K \left ( c_{k}  - \ao_k \right )^2 + \sum_{k=K+1}^{+\infty} c_{k}^2 \ .
\label{residual}
\end{align}

%  \ndp{Usare ${\bf \tilde{c}}$ al posto di $\ao$?}
 
A few observations can be made. In the case the true spectrum has finite power $\mathcal{P} = \int_{-\infty}^{+\infty} \So(\omega) \dd \omega$, the infinite sum $\sum_{k=K+1}^{+\infty} c_{k}^2$ is also finite. This term evaluates to the same value independently on how we extend the basis $\Fo_k$ for $k\geq K+1$. For this reason, we collect in the vector ${\bf c}$ only the entries $c_k,\ k=1, \ldots K$.

In some sense, the term $\sum_{k=K+1}^{+\infty} c_{k}^2$ is a residual error that stems from the fact that the finite set of filter function is not sufficient to represent the possibly infinite-dimensional spectrum. 
%the spectrum lives in an infinitesimal dimensional state and the number of filter functions $K$ is finite. \ndp{}
A good sensing procedure would thus choose a set of $F_n$ that keeps the main components $c_k$ of the true spectrum in the first term of \eqref{residual}, leaving smaller components in the second one. Therefore, the problem of minimizing the mean squared error with respect to ${\bf a}$ reduces to a LS minimization of the first term with respect to ${\bf \ao}={\bf B}^T{\bf a}$, that is, with respect to the coefficients of the estimated spectral density $\Sr(\omega)$ in the orthonormal basis $\{\Fo\}_{k=1}^{K}$, $\Sr(\omega) = \sum_{k=1}^K \ao_k \Fo_k(\omega)$.

To this end, note that in the orthonormal basis the vector of data $[{\boldsymbol \chi}]_n = \chi_n$ is written as ${\boldsymbol \chi} = {\bf B} {\bf c}$, since
\begin{align}
    \chi_n & = \int_{-\infty}^{+\infty} \So(\omega) F_n(\omega)\dd\omega = \int_{-\infty}^{+\infty} \So(\omega) \sum_{k=1}^K b_{n,k}\,\Fo_k(\omega)\dd\omega  \\
    & =\sum_{k=1}^K b_{n,k} \int_{-\infty}^{+\infty} \So(\omega) \Fo_k(\omega)\dd\omega = \sum_{k=1}^K b_{n,k}\,c_k \ .
\end{align}

Thus, we are allowed to rewrite the minimization problem of the first term of Eq.~\eqref{residual} as a function of the previously-defined matrices: 
\begin{align}
    \min_{\bf \ao} \quad  ({\bf c} - {\bf \ao})^T({\bf c} - {\bf \ao}) & = \min_{\bf a} \quad ({\bf c} - {\bf B}^T{\bf a})^T({\bf c} - {\bf B}^T{\bf a}) \label{def:min_a}\\
    & = \min_{\bf a} \quad  {\bf a}^T{\bf B} {\bf B}^T {\bf a} - {\bf c}^T{\bf B}^T{\bf a} -{\bf a}^T{\bf B}{\bf c} + {\bf c}^T{\bf c}   \\
    & ={\bf c}^T{\bf c} + \min_{\bf a} \quad  {\bf a}^T{\bf G}{\bf a} - 2 {\boldsymbol \chi}^T {\bf a} \ . 
\end{align}
Note that in \eqref{def:min_a}, we could just resolve by setting ${\bf c} = {\bf \ao}$, but both ${\bf c}$ and ${\bf a}$ are unknown, and for this reason we reformulate the problem in the known parameters ${\bf G}$, ${\boldsymbol \chi}$ and in the optimization variable ${\bf a}$.
By taking the singular value decomposition of the Gramian matrix ${\bf G} = {\bf U}{\bf \Lambda} {\bf U}^T$, we define ${\bf u}={\bf U}^T {\bf a}$, ${\bf x}={\bf U}^T {\boldsymbol \chi}$ and obtain
\beq
\min_{\bf u} \quad {\bf u}^T{\bf \Lambda}{\bf u} - 2 {\bf x}^T {\bf u} \ ,
\label{def:min_u}
\eeq
which has solution 
\beq
{\bf u}_k =\frac{{\bf x}_k}{{\bf \Lambda}_k},\ k=1, \ldots K, \quad \text{with} \quad {\bf a} = {\bf U}{\bf \Lambda}^{-1}{\bf U}^T{\boldsymbol \chi} \ .
\label{def:solution}
\eeq
The analytical solution \eqref{def:solution} gives the same spectral estimation of the Filter Orthogonalization procedure presented in Refs.~\citen{Mueller2018,MuellerSensingZeno2019}.
%\ndp{La soluzione è la stessa di \cite{Mueller2018} ponendo $G=A^{(M)},\ U^T=V^{(M)},\ \chi = c^{(M)}$ (indico con $^{(M)}$ le matrici del paper di Mueller), infatti
%\begin{align}
%    \tilde{S}^{(M)} & = \sum_{k=1}^K \tilde{c}_k^{(M)} \tilde{F}_k^{(M)} =  \sum_{k=1}^K \left(\frac{1}{\sqrt{\lambda_k}} \sum_{l=1}^K V_{k,l}^{(M)} c_l^{(M)} \right) \tilde{F}_k^{(M)} \\
%    & = \sum_{k=1}^K \left(\frac{1}{\sqrt{\lambda_k}} \sum_{l=1}^K V_{k,l}^{(M)} c_l^{(M)} \right) \left ( \frac{1}{\sqrt{\lambda_k}} \sum_{m=1}^K V_{km}^{(M)} F_m^{(M)} \right) \\
%    & =  \sum_{m=1}^K F_m^{(M)}  \sum_{k,l=1}^K \frac{1}{\lambda_k}  V_{km}^{(M)}  V_{k,l}^{(M)} c_l^{(M)} \\
%    & =  \sum_{m=1}^K F_m^{(M)}  \sum_{k,l=1}^K   U_{m,k} \frac{1}{\lambda_k} U_{l,k} \ \chi_l \\
%    & =  \sum_{m=1}^K F_m^{(M)} a_m, \quad a = U \Lambda^{-1} U^T \chi
%\end{align}
%che è la (9).}
Note that, in general, the rank of the Gramian matrix ${\bf G}$ is $K \leq N$. Sometimes it is beneficial to employ a truncated singular value decomposition\cite{Hansen1987}, that is, to consider only the highest $R$ eigenvalues $\lambda_k = {\bf \Lambda}_k, \ k=1,\ldots R$ and to approximate the Gramian with ${\bf G}^{(\rm trunc.)} = {\bf U}\,\textrm{diag}(\lambda_1, \ldots,\lambda_R, 0, \ldots,0) {\bf U}^T$.
%leave out the smaller ones. 
In this way, the components corresponding to the smallest eigenvalues, which are likely to be large and sensible to noisy data ${\boldsymbol \chi}$, are neglected, resulting in a smoother estimate.
%, or can be intentionally reduced by taking 
%thus, one is allowed just to take the highest $K$ eigenvalues $\lambda_k = {\bf \Lambda}_k$ and leaving out the smaller ones. This is equivalent to use a reduced Gramian matrix  ${\bf G}^{(\rm red)} = {\bf U}\,\textrm{diag}(\lambda_1, \ldots,\lambda_K, 0, \ldots,0) {\bf U}^T$ in place of ${\bf G}$, so as to avoid the ``explosion'' in magnitude of some components of ${\bf u}$ and ${\bf a}$ when ${\bf \Lambda}_k$ is too small.

As a concluding remark, we show that the solution obtained in Eq.\,\eqref{def:solution} coincides with the solution of Eq.\,\eqref{def:ls_system} obtained via pseudoinverse,
\beq
{\bf \So} = ({\bf F}^T{\bf F})^{-1}{\bf F}^T {\boldsymbol \chi}\ , 
\label{def:pseudoinverseSolution}
\eeq
when the filter function and the spectrum are discretized in the frequency domain with the same discretization $\Delta \omega$ that can be used to numerically evaluate the integral \eqref{def:gramian} defining the Gramian. 
The following is not meant to be a formal proof but to rather give an intuition of the equivalence.
In this respect, let us define ${\bf F}$ as a rectangular matrix with $N_R=N$ rows, with $N_R$ the number of controls, and $N_C$ columns, where $N_C$ denotes the number of unknown ${\bf S}$. As previously anticipated, usually $N_C \gg N_R$, with the consequence that the squared matrix ${\bf F}^T{\bf F}$ is rank deficient. The matrices ${\bf F},\ {\bf G}$ and ${\bf F}^T{\bf F}$ are related via their singular value decomposition, \beq
{\bf F} = {\bf U} \sqrt{{\bf \Lambda}} {\bf V}^T, \quad {\bf G} = {\bf F}{\bf F}^T={\bf U}{\bf \Lambda}{\bf U}^T, \quad {\bf F}^T{\bf F} = {\bf V}{\bf \Lambda}{\bf V}^T
\eeq
where ${\bf U},\, {\bf \Lambda},\ {\bf V}$ have been extended in their size but keeping the correct images and kernels, and verifying the relations ${\bf U}^T{\bf U} = {\bf I}$, ${\bf V}^T{\bf V} = {\bf I}$. Then, we get
\begin{align}
{\bf \So} & = ({\bf F}^T{\bf F})^{-1}{\bf F}^T {\boldsymbol \chi} = ({\bf V} {\bf \Lambda} {\bf V}^T)^{-1} {\bf V} \sqrt{\bf \Lambda} {\bf U}^T {\boldsymbol \chi} = {\bf V} (\sqrt{{\bf \Lambda}})^{-1} {\bf U}^T {\boldsymbol \chi} \\
& = {\bf V} \sqrt{{\bf \Lambda}} {\bf U}^T {\bf U} {\bf \Lambda}^{-1} {\bf U}^T {\boldsymbol \chi} = {\bf F}^{T}{\bf a} \ .
\end{align}
This means that the solution by pseudoinverse is a linear combination of filter functions given by the optimal coefficients {\bf a} of Eq.\,\eqref{def:solution}. The reason for a bad estimate via pseudo-inversion lies in the difficulty and numerical instability in finding the inverse 
%pseudo-inverse 
of ${\bf F}^T{\bf F}$.

%\ndp{${\bf F}^T{\bf F}$ has its own proper name, but I can't find it. Anyone remembers it?}
%
%\ndp{nota sul rango di $G, \lambda$ e il valore K, cosa succede se $\lambda_k =0$}
%\mmm{Is this the same result for $\bf a$ that we had in the previous paper? Then, we should strengthen in the discussion that we have gained new insight on the quality of the solution.}

\subsection{Non-negative Least Squares Estimation}
% Assuring the non-negativeness of the spectral estimate}

%\ndp{inserire vincolo sulla positività di c, le F sono positive? $a > 0$?}
%\ndp{se $F_n$ hanno supporto disgiunto otengo il vincolo di $a_n>0$}
%\mmm{this would be great. Can this be obtained from the approach above by using non-negative least squares?}

The minimization problems in Eqs.\,\eqref{def:min_a} and \eqref{def:min_u} are 
%appear as 
unconstrained problems in the optimization variables ${\bf a}$ and ${\bf u}$. 
%While this allows to find the solution \eqref{def:solution}, it leave behind the fact that a correct spectral estimate $\Sr(\omega)$ should verify $\Sr(\omega) \geq 0$ for all frequencies $\omega$. 
Being $\So(\omega)$ a power spectral density, a correct estimate should verify $\Sr(\omega) \geq 0$ for all frequencies $\omega$. 
However, due to errors in the measurement
%estimation 
of ${\bf \chi}$ or due to the finite-size subspace estimation of $\So(\omega)$, we may obtain a solution with $\Sr(\omega)<0$ for some frequencies $\omega$.

When this occurs, we may want to enforce our solution to give a non-negative estimate $\Sr(\omega)$.
%, that is, we want to solve the problem
This is ensured by solving the optimization problem
\begin{align}
& \min_{\bf a} \quad ({\bf c} - {\bf B}^T{\bf a})^T({\bf c} - {\bf B}^T{\bf a}) \label{def:qp_minimization}\\
& \textrm{subject to}\,\,\sum_{n=1}^N a_n F_n(\omega) \geq 0,\quad \forall \omega \in [0,+\infty) \, . \label{estimate_constaints}
\end{align}
If the filter functions $F_n(\omega)$ have disjoint support or the estimation error due to the overlap between filters is negligible when compared to other sources of errors, the constraints \eqref{estimate_constaints} reduces to $a_n \geq 0, n=1, \ldots N$, since any $a_n<0$ would give $\Sr(\omega)<0$ in the support of $F_n(\omega)$. With this assumption,
%In the following we assume that the filter functions $F_n(\omega)$ have disjoint support, or that their overlap is negligible so that this assumption is justified when compared to other sources of errors. With this assumption, the constraints \eqref{estimate_constaints} reduces to $a_n \geq 0, n=1, \ldots N$ since any $a_n<0$ would give $\Sr(\omega)<0$ in the support of $F_n(\omega)$. 
the problem can be formulated as a non-negative least squares problem, 
\beq
\min_{a_n \geq 0} \quad ({\bf c} - {\bf B}^T{\bf a})^T({\bf c} - {\bf B}^T{\bf a}) \, ,
\eeq
which with the appropriate substitutions becomes
\beq
\min_{a_n \geq 0} \quad  {\bf a}^T{\bf G}{\bf a} - 2 {\boldsymbol \chi}^T {\bf a} \, .
\label{def:NNLS}
\eeq
This is a quadratic programming problem whose solution can be found numerically by using standard routines.\cite{BoydBook} 
% with Interior Point methods \ndp{REFERENCE}.
%\mmm{if we can get a numerical example for this and see how the fidelity compares to the case where we would take the orthogonalization solution and simply set it to zero where negative values occur}

\section{\label{sec:design} Improved design of filter functions}

In the previous sections, we have assumed the filter functions to be given in the formulation of the problem. 
%In this section we propose an improvement for a set of filter functions used in a previous work\,\cite{Do2019}
Here, we propose a novel strategy for the improvement of their design, with a particular focus on the filter functions already used in Ref.~\refcite{Do2019}\,. While this is far from being a complete optimization of the control signals, we believe that this proposal can be applied to other sensing protocol to design high performing
%to improve the design of the 
filter functions. %The reader interested in a more detailed mathematical description of the connection between the dynamical decoupling sequence construction and filter-design theory can refer to Ref\cite{Biercuk2011}.

\begin{figure}[h]
\centering    
\subfigure[]{\label{fig:filter_fo}\includegraphics[width=0.49\textwidth]{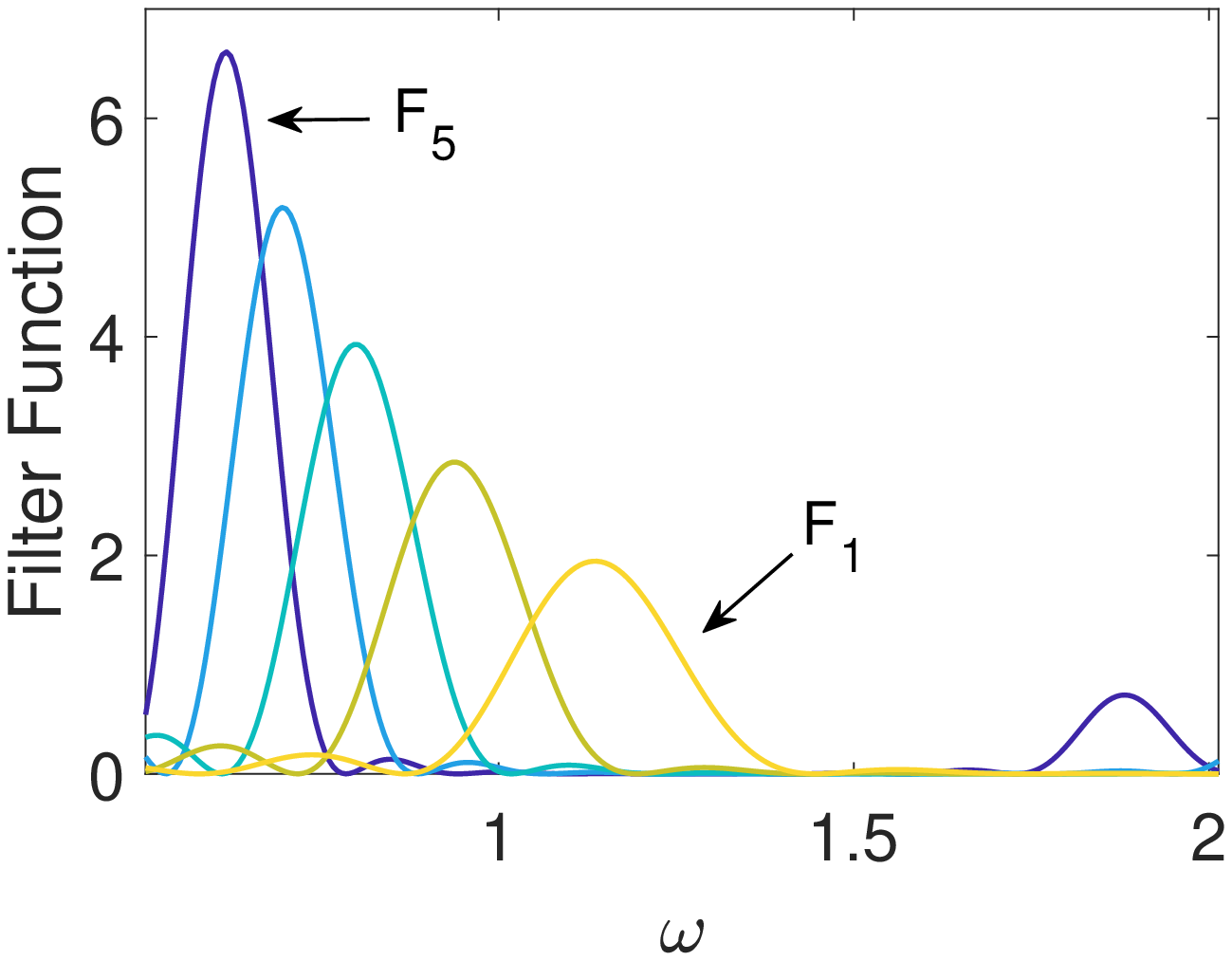}}
\subfigure[]{\label{fig:filter_equi}\includegraphics[width=0.49\textwidth]{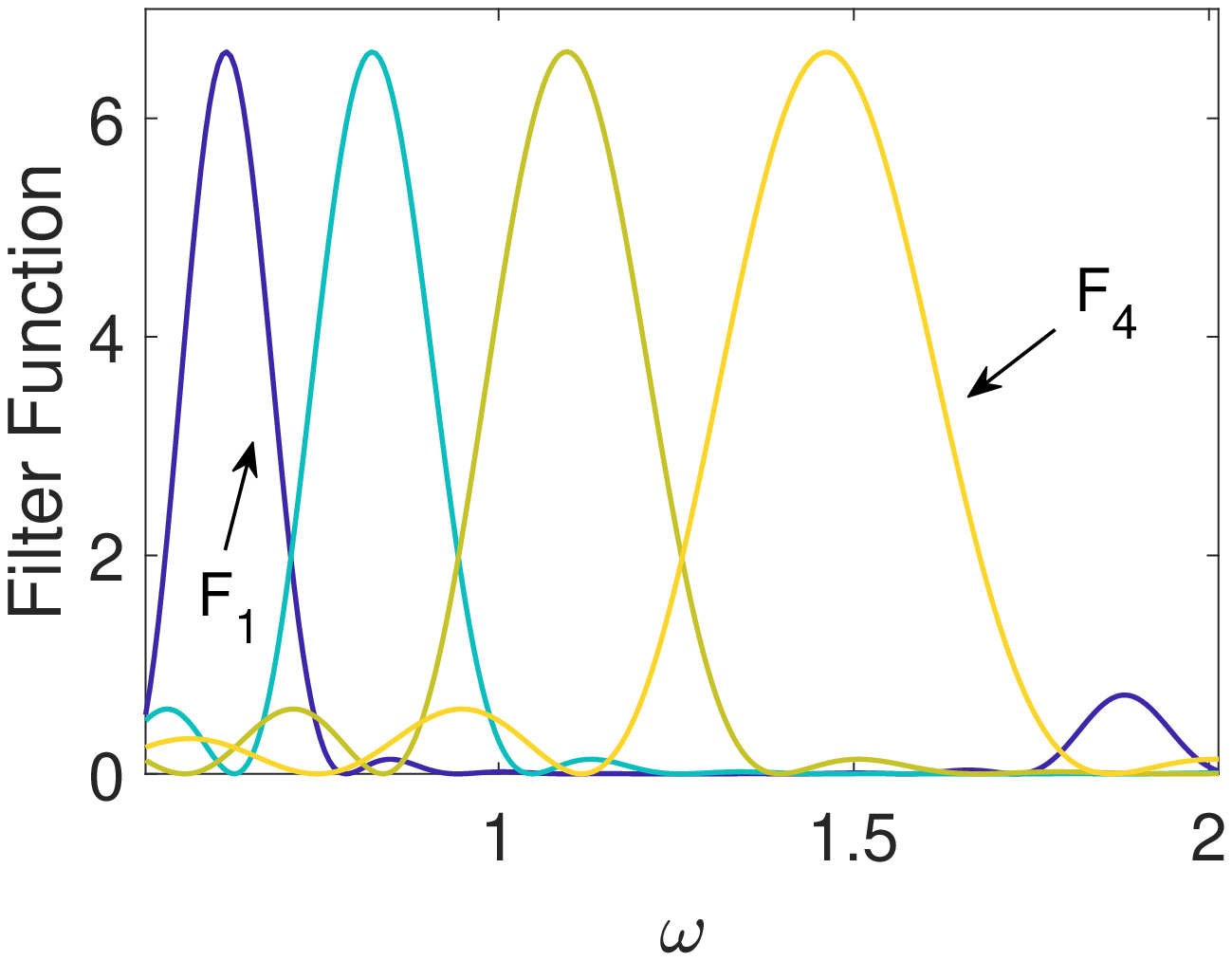}}
\caption{Different designs of filter functions $F_k$, expressed in [10$^{18}$ Hz$^2$], as a function of $\omega$ [$10^6 \times$ rad/sec]. %The scale on the axes refers to the numerical examples of Section \ref{sec:numerical}. 
(a) From right to left, the main peak of the filter functions $F_1$ (yellow) to $F_5$ (blue) with interpulse duration multiple of a minimum period, $\tau_n=n\tau_1$. The main peak of the filter function, corresponding to the first harmonic at $\frac{\pi}{\tau_n}$, accumulates on the lower frequencies.  Amplitudes scale as $\tau_n^2$. For $F_5$ one can also see the peak corresponding to the next harmonic (second blue smaller peak on the right corner). (b) Filter functions designed with fixed bandwidth overlap between main peaks, plotted from $F_1$ (blue, on the left) to $F_4$ (yellow, on the right). Note that the control amplitude $A_c^{(n)}$ has been scaled to have the same amplitude in the main peak (see main text).}
\label{fig:filter}
\end{figure}

The intuition behind this improvement comes from the attempt to solve the linear equation system in \eqref{def:ls_system}. As we have already anticipated, the inversion of the relations \eqref{chiN} would be easier and immediate if the filter functions excite few disjoint frequency components of the noise power spectral density. To verify if this occurs, let us analyze the frequency domain of a commonly-used filter function set. 
In this regard, we consider PDD sequences, \cite{Khodjasteh2005,Alvarez2011,Degen2017} that is, a family of multipulse sensing sequences with $M$ equally spaced sign flips of $\Omega_c$  at $t_j = j \tau,\ j=1,\ldots M$ (realized by $\pi$-pulses\cite{Degen2017} or Zeno projective measurements\cite{Do2019}) parametrized by the interpulse duration $\tau$. These sequences originate piecewise constant control signals $\Omega_c^{(n)}(t)$ that switch between opposite values $A_c^{(n)}$ at time instants $t_j$. Such control signals are easy to be generated and, thus, they are routinely implemented in experimental setups. In the formulation of the previous sections, the index $n=1, \ldots N$ refers to a filter function with specific duration $\tau_n$. 
A common choice for the interpulse duration is to employ the multiples of a fixed minimum value $\tau_n= n \tau_1$. While this choice seems natural, a plot of the filter functions reveals an imbalance in the range of the covered frequencies. In fact, the Fourier transform of the control $\Omega_c^{(n)}(t)$, which is composed by $M/2$ periodic square pulses, reads
\beq
\mathcal{F}[\Omega_c^{(n)}](\omega) = \displaystyle{A_c\,\tau_n\,2^{\log_{2}M}\ii \ee^{-\ii \frac{M\omega\tau_n}{2}} \textrm{Sinc} \left(\omega \frac{\tau_n}{2} \right) \sin \left(\omega \frac{\tau_n}{2} \right) \Pi_{k=0}^{\log_{2} M-2} \cos \left(2^k \omega \tau_n \right)} \ , 
\eeq
where $\textrm{Sinc}(\cdot)$ denotes the sinc function and $M$ is assumed to be a power of 2. The peaks of the corresponding filter function are at the odd harmonics $\omega_p=p \frac{\pi}{\tau_n}$, $ p=1,3,\ldots $ The main peak is the first one, with an amplitude of $4A_c^2M^2\tau_n^2/\pi^2$ that gives the greatest contribution to excite the corresponding frequency of the noise power spectral density.
Around each harmonic the filter function sharply goes to zero, becoming null for $\omega = \frac{\pi}{\tau_n} \pm \frac{2\pi}{M \tau_n}$. Since the position of the peak depends on $\tau_n$, 
%as $n$ increases the main peaks of the filter functions overlap more and more 
the overlap between the filter functions is more pronounced for an increasing value of $n$ and thus of $\tau_n$. Overall, the frequency interval covered by the main peaks of all filter functions ranges from $\omega_{\rm min} = \frac{\pi}{\tau_N}$ to $\omega_{\rm max} = \frac{\pi}{\tau_1}$, but is not evenly sampled since filter functions with higher $n$ overlaps in 
%their peaks bandwidth. 
the bandwidth spanned by their peaks. In Fig.\ref{fig:filter_fo} we plot the main peak of five consecutive filter functions. Both the spread of the peaks and the reduction in the amplitude are clearly visible. 

As a novel strategy, we propose to set first the amount of overlap in the peak bandwidth of the filter functions,
% to be implemented, 
and then to evaluate $\tau_n$. Specifically, let us consider to have
%one single 
a filter function whose main peak is
%that we have a filter function with the main peak 
at $\frac{\pi}{\tau_n}$.
%we first assign the lower frequency $\omega_{min}$ of the band that we want to investigate, therefore setting $\tau_n = \tau_{max} = \frac{\pi}{\omega_{min}}$.
The frequency range around the main peak goes up to $\left( 1+\frac{2}{M} \right)\frac{\pi}{\tau_n}$. Then, take a second filter function
%Considering now the next filter function defined by $\tau_{n+1}$, we define the lower bandwidth of its main peak to verify 
defined by $\tau_{n+1}$. The latter is chosen such that the lower bandwidth of the main peak is at the $1-\varepsilon$ fraction of the upper band of the previous filter function, characterized by a larger value of $\tau_n$. More formally, $\tau_{n+1}$ has to verify the following relation:
\beq
\left( 1 - \frac{2}{M}\right)\frac{\pi}{\tau_{n+1}} = \left( 1+\frac{2}{M} -\varepsilon\frac{4}{M}\right) \frac{\pi}{\tau_n} \ .
\label{def:bandwidthRelation}
\eeq
%that is, to be at the $1-\varepsilon$ fraction of the upper band of the previous filter function. 
The parameter $\varepsilon \in [0,1]$ hence defines the bandwidth overlap of the main peaks of consecutive filter functions. For $\varepsilon=0$ the peaks have disjoint support, for $\varepsilon=1$ the peaks completely overlap and the filter functions coincide. Solving \eqref{def:bandwidthRelation} for $\tau_{n+1}$ gives
\beq
\tau_{n+1} = \left(\frac{M-2}{M+2-4\varepsilon} \right)\tau_n = \left(\frac{M-2}{M+2-4\varepsilon} \right)^n \tau_{1} \ ,
\label{eq:recursiveTau}
\eeq
%\beq
%\tau_{n+1} = \left (1 - \frac{4 (1 - \varepsilon)}{M + 2 - 4 \varepsilon} \right) \tau_n = \left (1 - \frac{4 (1 - \varepsilon)}{M + 2 - 4 \varepsilon} \right)^n \tau_{max},
%\label{eq:recursiveTau}
%\eeq
which can be used to define $\tau_{n}$ starting from $\tau_1$ in decreasing order. The recursion is stopped when the last considered filter function overlaps the next not null harmonic 
%We decide to stop when the next filter function would overlap the next harmonic 
of the first filter function, which is placed at $\omega_{max}=\frac{3\pi}{\tau_{1}}$. In the case this overlap does not affect the estimation fidelity, in principle we can continue to evaluate $t_n$ up to a subsequent not null harmonic. For instance, in Section \ref{sec:numerical} we consider the set BOD(3) of filter functions designed up to the 3rd harmonic of $F_1$ and the set BOD(5) designed up to its 5th harmonic.

To summarize, the design procedure of the filter function is the following:
\begin{itemize}
    \item[1 -] From the bandwidth $[\omega_{\rm min}, \omega_{\rm max}]$ 
    %that we want to investigate, 
    to be investigated, define $\tau_{\rm max}=\tau_1$ such that $\frac{\pi}{\tau_{1}}<\omega_{\rm min}$ and $\omega_{\rm max}<\frac{3\pi}{\tau_{1}}$
    \item[2 -] Set the number $N$ of sample points in $[\omega_{\rm min}, \omega_{\rm max}]$, the overlap $\varepsilon$ and $M$ (must be a power of 2).
    %to be estimated 
    %choose $M$ as the next power of 2 of the number of sample points $N$ that you want to estimate in $[\omega_{\rm min}, \omega_{\rm max}]$. 
    Since $\tau_N \approx \tau_{1}/3$, a rough estimate of $N$ can be obtained solving Eq.\,\eqref{eq:recursiveTau} with $n=N-1$ as a function of $M,\ \varepsilon$.
    \item[3 -] Use Eq. \eqref{eq:recursiveTau} to evaluate  $\tau_{n+1}$, with $n=1,\ldots N-1$ starting from $\tau_{1}$
    \item[4 -] Optionally, the amplitude of the control signals in the Zeno sensing protocol can be adjusted to have the same amplitude of the main peak by putting $A_c^{(n)}~=~A_c^{(1)} \tau_1/\tau_n $. %\frac{\tau_1}{\tau_n}
\end{itemize}

\section{\label{sec:numerical}Numerical Simulations}

In this section we show the results of the numerical simulations that we have performed to compare different sensing protocols. We make the comparison between two sensing protocols with the one obtained using the bandwidth-overlap filter functions Design, both with LS and NNLS estimation.

Specifically, we consider as control signals $\Omega_c(t)$ both PDD sequences, with $M$ equally spaced sign flips %$\pi$-pulses 
at $t_j = j \tau,\ j=1,\ldots M$, and CP multipulse controls,\cite{Carr1954,Alvarez2011} whereby the sign flips occur at $t_j=\frac{2j-1}{2}\tau$. Both these sequences are squared waves with zero mean parametrized by the interpulse duration $\tau$.
Note that, if we consider the same scenario of the experimental setup of Ref.~\refcite{Do2019}, a projective measurement is performed at every time instant $t_j$.
%before the application of a control $\pi$-pulse. 
Each projective measurement rapidly brings the atoms into the ground state of the quantum probe with a given probability, thus allowing to keep the system in the weak Zeno regime.\cite{Mueller2016b,Mueller2018}
%, which can be leveraged to enhance the sensitivity of the probe
%However, while a PDD sequences is eligible to be a control pulse for this sensing method, the same does not longer hold for the CP pulses. Indeed, each CP pulse would be applied in the middle of the time interval between two consecutive projective measurements, with the negative consequence that the effects of control do not change the atomic population in the quantum probe ground state. 
This allows to recover the universal formula \eqref{universalRelation}, which is the base to set the estimation problem discussed in this paper.

In the simulations we consider sequences of 32 sign flips of $\Omega_c$ (realized by $\pi$-pulses\cite{Degen2017} or Zeno projective measurements\cite{Do2019}). For the CP and PDD protocols, we set up 32 control sequences. Their interpulse duration $\tau_n$ linearly space from $t_1 = 1 \mu s$ to $t_{32} = 5 \mu s$, corresponding to the frequency interval ranging from $\omega_{\rm min} = \frac{\pi}{\tau_{32}}=6.28\times10^5Hz$ to $\omega_{\rm max} = \frac{\pi}{\tau_{1}}=3.14 \times10^6Hz$. 
%For the estimate of the noise power spectral density, we set up the unconstrained optimization problem and use the solution \eqref{def:solution} with these protocols. 
Instead, in the case of the BOD protocol, we choose $\tau_1=5 \mu s$ and evaluate $t_n$ with $\varepsilon = 0.5$ for two sets of filter functions, BOD(3) and BOD(5), the former with 17 
%filters 
control pulses (up the third harmonic of $F_1$, $\tau_{17}=1.8 \mu s$), and the latter with 25 filters (up to the fifth harmonic of $F_1$, $\tau_{25}=1.1 \mu s$). The amplitudes of the control signals are chosen to verify the conditions of the weak Zeno regime.\cite{Mueller2016b,Mueller2018,Do2019} %For the BOD protocol, we use the non-negative least square estimate to solve the estimation problem.

\begin{figure}[h]
\centering    
\includegraphics[width=0.7\textwidth]{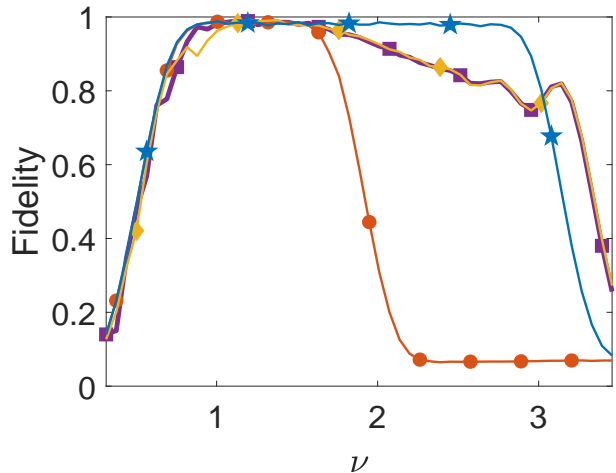}
\caption{Reconstruction fidelity by using the protocols PDD (purple line, square markers), CP (yellow line, diamond markers), BOD(3) (red line, circle markers) and BOD(5) (blue line, star markers) when estimating a Gaussian power spectral density with central frequency $\nu$ [$10^6~\times$~rad/sec].}
\label{fig:scan}
\end{figure}

\begin{figure}[h]
\centering    
\subfigure[PDD]{\label{multi_fo}\includegraphics[width=0.49\textwidth]{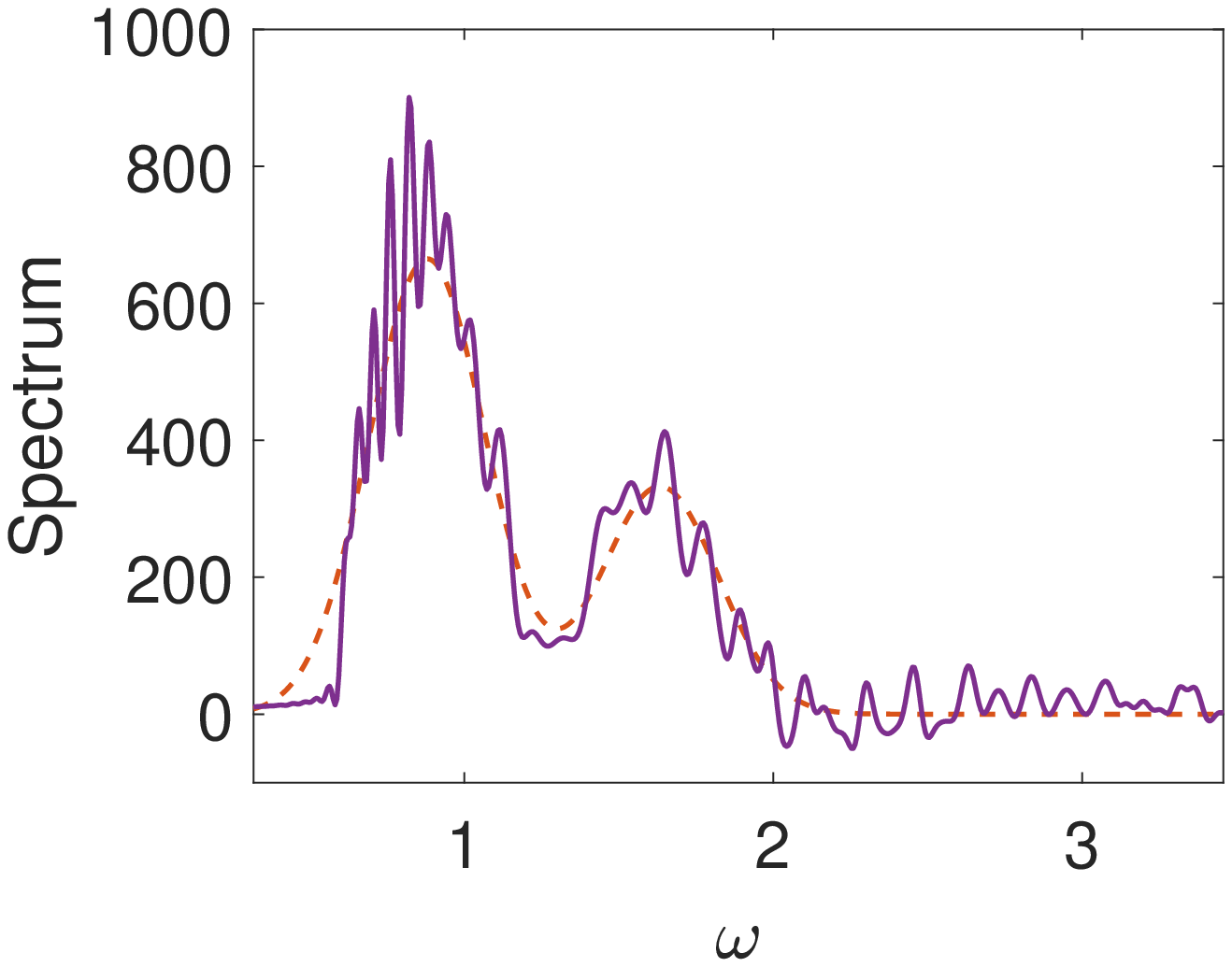}}
\subfigure[CP]{\label{multi_as}\includegraphics[width=0.49\textwidth]{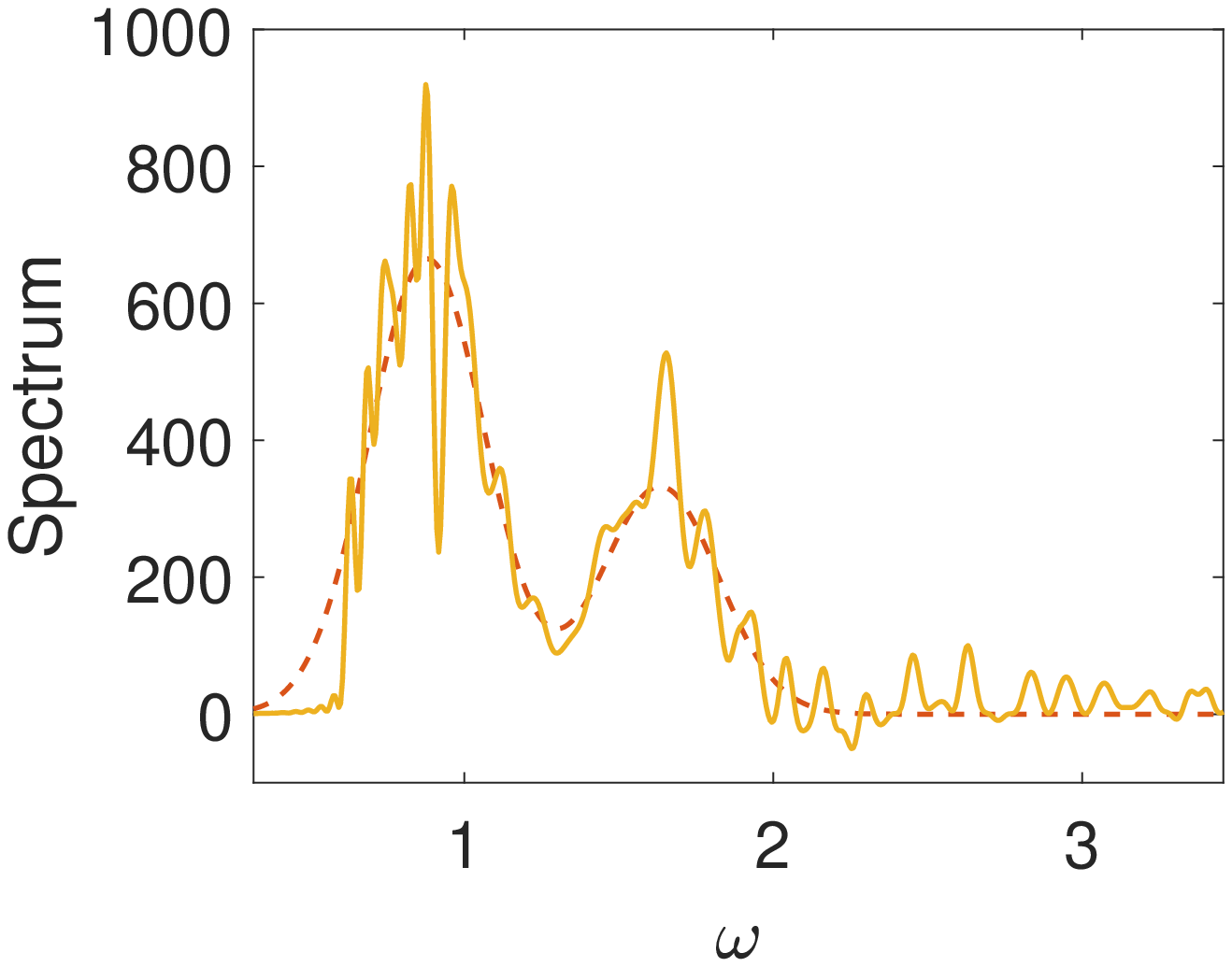}}\\
\subfigure[BOD(3)]{\label{multi_nnls}\includegraphics[width=0.49\textwidth]{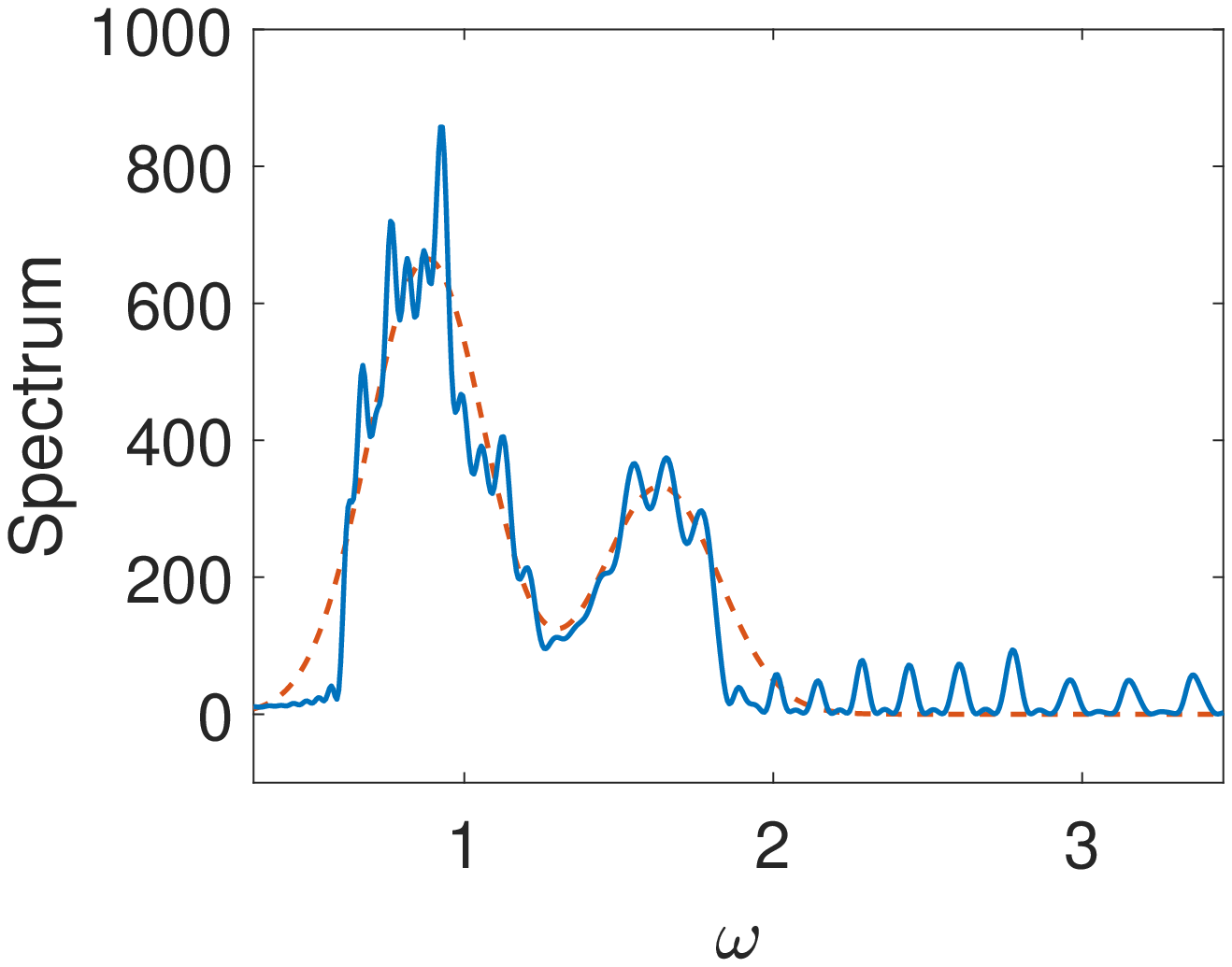}}
\subfigure[Pseudoinverse]{\label{multi_ls}\includegraphics[width=0.49\textwidth]{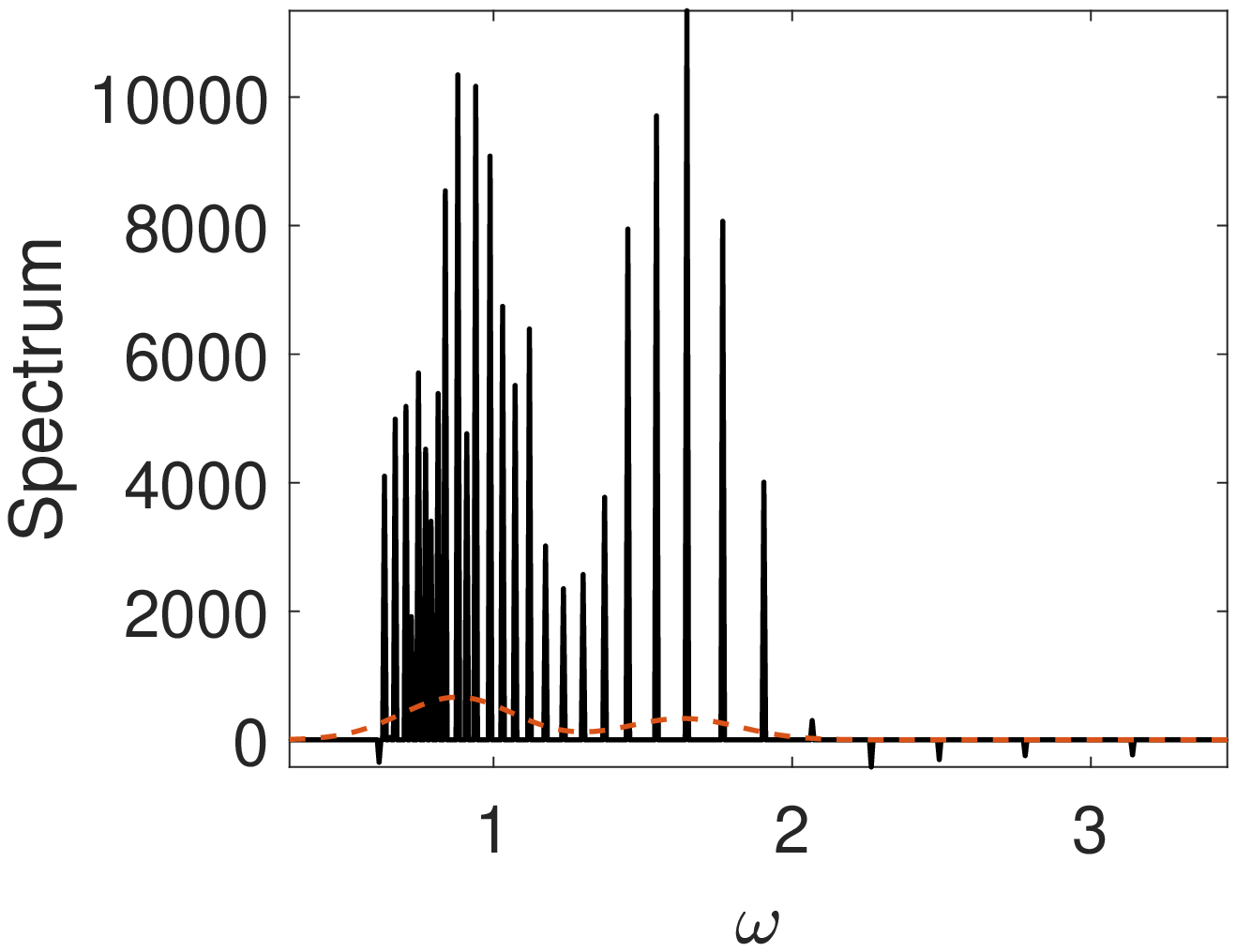}}
\caption{Qualitative evaluation of the sensing protocols performance by comparing the obtained estimates $\Sr(\omega)$, $\omega$ in $10^6 \times$ rad/sec. The protocols PDD, CP and BOD(3) use the corresponding sequences described in the main text and the LS estimate provided by Eq.\,\eqref{def:solution}. The pseudoinverse method refers to the estimation obtained directly by Eq.\,\eqref{def:pseudoinverseSolution}, sampling the spectrum and the filter functions with $\Delta \omega = 6\times10^3 Hz$. }
\label{fig:multi}
\end{figure}

%Two tests are performed on these protocols. 
As a first test on these protocols, we consider a Gaussian power spectral density, i.e.
\beq
\So(\omega) = \frac{\mathcal{N}}{2\sqrt{2\pi \sigma^2}}\ee^{-\frac{(\omega-\nu)^2}{2\sigma^2}}, \quad \omega \geq 0,
\label{def:gaussianSpectrum}
\eeq
with $\mathcal{N} = 10^8 \ Hz^2,\ \sigma = 2\pi \times 30\,kHz$. 
%\ndp{$\mathcal{N}$ è una potenza, ma andrebbe riferita all'unità di misura reale di $\Omega(t)$. Teniamo Watt?}
We run the simulations for $\nu$ ranging from $2\pi \times 50\,kHz$ to $2\pi\times 550\,kHz$ every $10\,kHz$. Each spectral density is reconstructed for each protocol CP, PDD, BOD(3) and BOD(5) by using LS estimation. We calculate the fidelity between the true spectral density $\So(\omega)$ and the reconstructed one $\Sr(\omega)$ by means of the following relation: 
\beq
{\rm Fidelity} = \frac{\int_{-\infty}^{+\infty} \So(\omega)\Sr(\omega) \dd \omega}{\int_{-\infty}^{+\infty} \So(\omega) \dd \omega\int_{-\infty}^{+\infty} \Sr(\omega) \dd \omega} \ ,
\label{def:fidelity}
\eeq
which allows to evaluate the quality of the estimation up to a scale factor.
The results are shown in Fig. \ref{fig:scan}. In the plot, we can notice that the four lines corresponding to each protocol drop to zero outside the estimated frequency range. %This means that high-performance of the protocols can be achieved but just in specific frequency ranges, which is why in the beginning (Section \ref{sec:introduction}) we said that some prior knowledge of the spectrum helps. 
All these protocols perform well in the first half of the chosen range of $\nu$, up to about $2\pi \times 160kHz$. Then, the fidelity of BOD(3) quickly drops to zero since it has no filter functions with main peaks above this frequency. CP and PDD perform similarly, slowly dropping in performance in the range $(2\pi \times 160\,kHz, 2\pi\times 300\,kHz)$. Instead, on the same interval the protocol BOD(5) keeps the fidelity close to 1.
\\
As a second test, we face the estimation of a spectral density with two Gaussian components, i.e.
\beq
\So(\omega) = \frac{\mathcal{N}_1}{2\sqrt{2\pi \sigma_1^2}}\ee^{-\frac{(\omega-\nu_1)^2}{2\sigma_1^2}} + \frac{\mathcal{N}_2}{2\sqrt{2\pi \sigma_2^2}}\ee^{-\frac{(\omega-\nu_2)^2}{2\sigma_2^2}}, \quad \omega \geq 0,
\eeq
with $\mathcal{N}_1 = 10^8 Hz^2,\ \mathcal{N}_2 = 5\times10^7 Hz^2,\ \nu_1 = 2\pi\times 140\,kHz,\ \nu_2 = 2\pi\times 260\,kHz,\ \sigma_1 = \sigma_2 = 2\pi \times 30\,kHz$. This spectrum is plotted in red dashed line in Fig. \ref{fig:multi}.
%This original spectrum is represented in red dashed line . 
%We use 50 noise realization to evaluate $\chi_n$ for each filter function $F_n$.
\\
We consider the protocols PDD, CP and BOD(3) with LS estimation obtained by truncated singular value decomposition removing the 3 smallest eigenvalues. %We also consider the pseudoinverse estimation obtained directly by Eq.\,\eqref{def:pseudoinverseSolution}.
We plot the estimated spectrum within the frequency range spaced by the main peaks of the filter functions $F_n$ in Fig.\,\ref{fig:multi}. For each of the filter functions we use 50 noise realizations to evaluate $\chi_n$.

The protocols PDD and CP behave similarly, managing to locate the two Gaussian shapes in the power spectral density. However, in the right half of the estimated spectrum there are oscillations that brings the estimate to negative values. This is probably due to the fact that in the chosen frequency interval the filter functions are more sparse. The protocol BOD(3), instead, shows less oscillations in the estimate $\Sr(\omega)$. For comparison, we also plot the estimation that we obtain by the direct pseudo-inversion of \eqref{def:ls_system}, which is not accurate and unavoidably shows some negative values.

Finally, we compare the performance of the sensing protocols when using the LS optimization or the NNLS minimization. We consider different setups, with a small, medium, large dataset to evaluate $\chi_n$ as a sample mean from respectively 10, 50 or 200 samples. We compare the averaged fidelity over 250 simulations for the protocols PDD, CP, BOD(3) with overlap parameter $\varepsilon \in \{0.25,\ 0.5,\ 0.75\}$. The results are reported in Table~\ref{tab:fidelity}. In the LS estimation, we use a truncated singular value decomposition with the highest half of the Gramian matrix eigenvalues. Furthermore, we set to zero the negative components of the reconstructed spectrum before evaluating the fidelity.

\begin{table}
\centering
  \begin{tabular}{|c|c|c|c|c|c|c|c|}
    \hline
    \multicolumn{2}{|c|}{} &
        \multicolumn{2}{c|}{10 samples} &
            \multicolumn{2}{c|}{50 samples} &
                \multicolumn{2}{c|}{200 samples} \\
    Protocol & N & LS & NNLS & LS & NNLS & LS & NNLS \\
    \hline
    PDD & 32 & 0.8202 & 0.9129 & 0.8873 & 0.9664 & 0.9038 & 0.9849 \\
    CP & 32 & 0.8214 & 0.9117 & 0.8898 & 0.9657 & 0.9074 & 0.9844 \\
    BOD$_3\ (\varepsilon = 0.75)$ & 34 & 0.9533 & 0.9232 & 0.9890 & 0.9715 & 0.9970 & 0.9880 \\
    BOD$_3\ (\varepsilon = 0.50)$ & 17 & 0.8969 & 0.9200  & 0.9198 & 0.9795 & 0.9241 & 0.9926 \\
    BOD$_3\ (\varepsilon = 0.25)$ & 11 & 0.6252 & 0.8415 & 0.6386 & 0.8833 & 0.6418 & 0.8926 \\
    \hline
  \end{tabular}
\caption{Average fidelity of the reconstructed power spectral density as a function of the control protocol employed (PDD, CP or BOD(3)) and as a function of the noise realizations (samples) used to evaluate ${\boldsymbol \chi}$, with the LS and with NNLS approaches. Next to the protocol column, the number $N$ of filter functions employed. The fidelity values have been averaged over 250 simulations.}
\label{tab:fidelity}  
\end{table}

As we can see, the PDD and CP protocols perform similarly. Their averaged fidelity increases with the number of samples, i.e., with increasing precision of the measurement of $\chi_n$, and we can see an improvement when NNLS is employed. The performance of BOD(3) varies depending on the overlap parameter $\varepsilon$, showing higher fidelities with respect to PDD and CP for $\varepsilon \in \{0.5,\ 0.75\}$ and lower performances for $\varepsilon = 0.25$. An explanation for this behaviour could be that in the latter case the protocols have too few filter functions to provide a good estimation. Also, while for $\varepsilon \in \{0.25,\ 0.5\}$ the NNLS gives an improvement in the fidelity, in the case of $\varepsilon = 0.75$ the performance worsens slightly. A reason for this could be that in this case the filter functions overlap too much, and the constraints ${\bf a}_n\geq 0$ is too restrictive respect to \eqref{estimate_constaints}, such that the minimization \eqref{def:NNLS} gives a suboptimal solution compared to the minimization \eqref{def:qp_minimization}-\eqref{estimate_constaints}.

\section{\label{sec:discussion}Discussion and Conclusions}

In this paper we provide a deeper understanding about why the filter orthogonalization procedure works in estimating a power spectral density of noise. In fact, the method is equivalent to a least squares problem that minimize the MSE between the true spectral density and its reconstruction, obtained as a linear combination of orthogonalized filter functions. A set of linear independent filter functions gives a full rank Gramian matrix and therefore a good estimation, since each filter function adds new useful information to be employed in the estimation procedure. 

However, the least squares minimization may sometimes give unfeasible estimation when the reconstructed power spectral density is negative at certain frequencies. Hence, we have proposed a solution for this issue by formulating the MSE minimization as a NNLS optimization, i.e., by forcing the coefficients of the filter function combination to be non-negative. 

Framing the solution of the estimation problem as a linear combination of the filter functions allows also to evaluate the frequencies where the noise spectrum can be reconstructed. The latter turns to be useful to design the set of filter functions so that they uniformly cover the frequency interval of the power spectral density that we want to estimate. With this purpose, we have proposed a design, BOD, such that consecutive controls have a fixed overlap ratio between their main peaks.

We have tested the BOD protocol and the NNLS estimation against the PDD and the CP protocols over a set of significant parameters, such as the number of data samples and the overlap ratio, obtaining a range of values where each protocols gives the best performance.

Future investigations will address the possibility to further increase the estimated frequency interval, and possibly test the BOD protocol in experimental setups.

% abbiamo capito perchè la filter orthogonalization funziona, perchè è equivalente alla soluzione che ci restituisce un problema minimi quadrati. Se il problema ai minimi quadrati è ben posto, allora la soluzione è unica. Questo sposta il problema originario ad un problema più semplice legato allo scegliere un set di filter function linearmente indipendenti ed in numero pari ai samples della spectral density di noise da ricostruire. In tal caso, il set delle filter functions è detto completo ed il metodo FO rende tale set anche informativo. Tale condizione equivale ad avere che il rango del Gramiano sia pieno/full.

%dire che è stato risolto il problema di noise sensing sopra formulato introducendo opportuni vincoli dedutti da considerazioni fisiche. Questo ha permesso ...

% Filter function design: proponiamo un nuovo set di filter functions che hanno fra di loro lo stesso overlap (il rapporto della banda delle funzioni di filtro è fissato) nel dominio della frequenza. Questo metodo permette di risolvere il problema di inversione in maniera più efficiente a parità del numero di filter funciotn usate (vedi tabella, risultati numerici). 

\section*{Acknowledgments}

S.G., N.D.P., and F.C. were financially supported from the Fondazione CR Firenze through the project Q-BIOSCAN, PATHOS EU H2020 FET-OPEN grant no. 828946, and UNIFI grant Q-CODYCES. M.M. acknowledges funding from the EC H2020 grant 820394 (ASTERIQS).

% \bibliographystyle{ws-ijqi}
% \bibliography{sample}

\end{document}